\documentclass[prd,preprint,nofootinbib]{revtex4}

\usepackage{amsmath}
\usepackage{amsfonts}
\usepackage{amssymb}
\usepackage{color,graphicx,shortvrb,epsfig}
\usepackage{latexsym}
\usepackage[a4paper]{geometry} 
\usepackage{hyperref}
\usepackage{fullpage}

\renewcommand{\Im}{{\mathcal Im}}
\renewcommand{\Re}{{\mathcal Re}}
\newcommand{\wt}{\widetilde}
\renewcommand{\t}{\theta}

\begin{document}

\preprint{\vbox{\hbox{}\hbox{}\hbox{}
    \hbox{WIS/24/04-OCT-DPP} \hbox{hep-ph/0410319}}}

\vspace*{2.5cm}

\title{Phenomenological Consequences of Soft Leptogenesis}
\author{Tamar Kashti}
\email{tamar.kashti@weizmann.ac.il} \affiliation{Department of
Particle Physics,  Weizmann Institute of Science, Rehovot 76100,
Israel}

\begin{abstract}
Soft supersymmetry breaking terms involving heavy singlet sneutrinos
can be the dominant source of leptogenesis. The relevant range of
parameters is different from standard leptogenesis: a lighter
Majorana mass, $M\lesssim 10^9 $ GeV (allowing a solution of the
gravitino problem), and smaller Yukawa couplings, $Y_N\lesssim
10^{-4}$. We investigate whether the various couplings of the
singlet sneutrinos, which are constrained by the requirement of
successful `soft leptogenesis', can have observable phenomenological
consequences. Specifically, we calculate the contributions of the
relevant soft supersymmetric breaking terms to the electric dipole
moments of the charged leptons and to lepton flavor violating
decays. Our result is that these contributions are small.
\end{abstract}

\maketitle

\section{Introduction}
The discovery of neutrino masses and mixing makes the see-saw
mechanism \cite{Minkowski:1977sc,Gell-Mann:vs, Yanagida,
Mohapatra:1979ia} highly attractive. The existence of heavy
singlet neutrinos with Majorana masses $M$ and Yukawa couplings
$Y_N$ to the active neutrinos becomes very likely. This framework
can also dynamically generate the baryon asymmetry of the universe
by Leptogenesis \cite{Fukugita:1986hr}. The supersymmetric
extension of this model (SSM+N) is well motivated because it
protects the Higgs mass from large radiative corrections involving
the heavy neutrino. Supersymmetry must be broken, hence the SSM+N
includes soft supersymmetry breaking terms for the SSM fields and
for the singlet sneutrino field $\wt N$. The singlet sneutrino
soft terms, in turn, can affect leptogenesis and even give the
dominant contribution for low enough masses ($M\lesssim 10^9$ GeV)
and small Yukawa couplings ($Y_N\lesssim 10^{-4}$)
\cite{Grossman:2003jv, D'Ambrosio:2003wy, Grossman:2004dz}.

In general, only the lightest singlet (s)neutrino contributes to
leptogenesis, since its interactions wash out any lepton asymmetry
created by the heavier singlet sneutrinos. However, this lightest
singlet (s)neutrino, even with $M\sim \mathcal{O}(1 \mbox{ TeV})$,
is practically impossible to discover at colliders, since the Yukawa
couplings need to be small enough in order for the decay to be out
of equilibrium. Singlet (s)neutrinos contribute, however, to
electric dipole moments (EDM) of leptons, $d_{\ell}$, and to lepton
flavor violating (LFV) interactions
\cite{Minkowski:1977sc, Hisano:1996qq, Casas:2001dv,Masina:2002mv, Masina:2003wt,%
Arganda:2004bz, Masiero:2004js}. These measurements already
constrain various SSM soft breaking terms
\cite{Ibrahim:1998je,Ellis:2001yz}, especially their phases which
should be $<\mathcal{O}(10^{-2})$ to be consistent with the non
observation of EDMs \cite{Abel:2001vy} (however, this limit is
significantly weakened if $m_{SUSY}\sim 1\mbox{ TeV}$
\cite{Kizukuri:1991mb}). Much work has been devoted to the
possible connection between leptogenesis and EDM
and LFV measurements \cite{Ellis:2002eh, Ellis:2002xg, Pascoli:2003uh,%
Dutta:2003my, Tobe:2003nx, Davidson:2003cq,  Raidal:2004vt}.

In this paper, we briefly describe the SSM+N model in section
\ref{NOTESsec}. Then we estimate the contribution of soft breaking
terms that can induce leptogenesis to $d_\ell$ in section
\ref{EDMsec}, and to the LFV interactions $Br(\ell_i\to
\ell_j\gamma)$ and $R(\mu\to e \mbox{ in Ti})$, in section
\ref{LFVsec}. We conclude in section \ref{CONCsec}. Our main
result is that soft leptogenesis gives small low-energy
phenomenological effects, so that present and near future
experiments of EDM and LFV do not constrain it.

In our previous works \cite{Grossman:2003jv,Grossman:2004dz}, we
considered for simplicity a one generation model. Since soft
leptogenesis constrains only the couplings of the lightest
sneutrino, $\wt N_1$, we only consider its contribution to the EDM
and to LFV interactions and not all $\wt N_i$ \nopagebreak[4]
contributions.
\section{The Model}
\label{NOTESsec} The relevant superpotential terms of the  SSM+N are
\begin{align}\label{SuperP}
W=Y_N LH_u \bar N+M \bar N \bar N +Y_L LH_d \bar E +\mu H_u H_d\;,
\end{align}
where $L$ is the supermultiplet containing the left handed lepton,
$N,E$ are the SU(2)-singlet superfields of the neutrino and
charged lepton respectively, and $H_u, H_d$ are the Higgs
superfields. The relevant soft breaking terms are
\begin{align}\label{SSB}
    -\mathcal{L}_{soft}=&B \wt N \wt N+m_2 \wt W^a\wt W^a +m_{\wt L}^2 \wt L^\dag \wt L
    +A_N \wt L H_u \wt N^\dag +A_L \wt L H_d \wt E^\dag
     +\text{H.c.}\;.
\end{align}
Here $\wt W^a$ ($a=1,2,3$) are the $SU(2)_L$ gauginos, $\widetilde
N,\widetilde L,H_u,H_d$ are scalar fields (and $N,L,h_u,h_d$ are
their fermionic superpartners). For a one generation model the
Lagrangian derived from eqs.~(\ref{SuperP}) and (\ref{SSB}) has
three independent physical CP violating phases:
\begin{align}\label{phases}
 \phi_N&=\arg(A_N Y_N^* MB^*)\;,\cr
 \phi_W&=\arg(m_2 Y_N A_N^*)\;,\cr
 \phi_L&=\arg(A_L Y_L^* MB^*)\;.
\end{align}
After spontaneous symmetry breaking, another phase is added:
\begin{align}\label{SSB phase}
    \theta&=\arg(m_2\mu v_u v_d)\;,
\end{align}
where $v_i$, for $i=u,d$ are the vacuum expectation values of the
Higgses, which are complex in general. In
\cite{Grossman:2003jv,Grossman:2004dz} we investigated the
contributions of $\phi_N$ and $\phi_W$ to leptogenesis.

\section{The EDM of the Electron}
\label{EDMsec} The CP violation in lepton interactions, that is
necessary to generate leptogenesis, is likely to induce electric
dipole moments for the charged leptons (for a review on EDMs, see
\cite{Bernreuther:1990jx}). We estimate the size of these
contributions, and compare them to experiment results. The present
bounds on the EDM of charged leptons are currently
\begin{align}\label{EDM exp}
    d_e &< 1.5\times 10^{-27} \mbox{e cm} \mbox{ \cite{Commins:gv}}\;,\cr
    d_\mu &< 2.8\times 10^{-19} \mbox{e cm} \mbox{ \cite{Deng:2003}}\;,\cr
    d_\tau &< 3\times 10^{-16} \mbox{e cm} \mbox{ \cite{Hagiwara:2002fs}}\;.
\end{align}
The near future expected sensitivities are
\begin{align}\label{EDM future}
    d_e < 10^{-33} \mbox{e cm} \mbox{ \cite{Lamoreaux:2001hb}}\;,\cr
    d_\mu < 10^{-26} \mbox{e cm} \mbox{ \cite{Aysto:2001zs}}\;.
\end{align}

We use here the mass eigenstate formulas of \cite{Ecker:1983dj} to
calculate the one loop contribution to the electron EDM using the
mass basis of the fields. Given an interaction of a lepton $\ell$
with a sneutrino $\wt\nu$ and a chargino $\chi$ of the form
\begin{equation}\label{C LR definations}
\mathcal{L}=-\sum_{ija}\bar{\ell}(C^x_{L\ell a}P_L+C^x_{R\ell
a}P_R)\chi_a \wt\nu_x + \text{h.c.}\;,
\end{equation}
where $\ell=e,\mu,\tau$, one obtains the following one-loop
contribution to $d_e$:
 \begin{equation}\label{CC in EDM}
d_e=\frac{m_\chi}{16\pi^2 m_{\tilde\nu_x}^2}I_4
\left(\frac{m_{\chi_a}^2}{m_{\tilde\nu_x}^2},\frac{m_e^2}{m_{\tilde\nu_x}^2}\right)
\Im(C^x_{L e a}C^{x*}_{R e a})\;,
\end{equation}
where
\begin{equation}\label{I4}
I_4(r,s)=\int_0^1dx\frac{x^2}{1-x-rx-sx(1-x)}\;.
\end{equation}

\begin{figure}[tb]
    \includegraphics[width=1\textwidth]{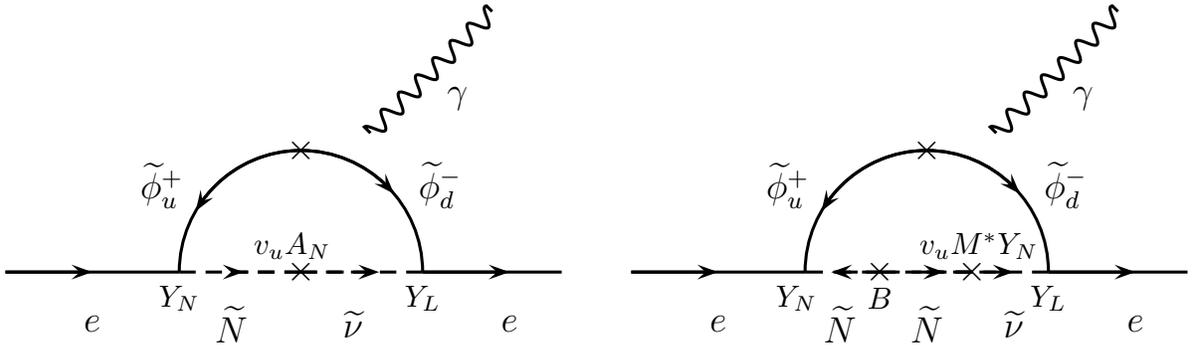}
    \caption{\it The two leading contributions to the EDM of the electron from sneutrino mixing.
    \label{EDM fig}}
  \end{figure}
The two leading contributions that involve the interactions of the
singlet sneutrinos come from the diagrams of fig.~\ref{EDM fig}.
In the sneutrino mass matrix, we neglect terms of order $|v_u\mu
Y_N/M^2|$, assume the hierarchy $|v_u A_N|, |v_u MY_N|\ll |B|\ll
|M|^2$, and write the sneutrino mass eigenstates up to order
$|Bv_u(A_N\pm MY_N)/M^4|$. We obtain:
\begin{align}\label{EDM th result}
&d_e\cong\frac{Q_e |m_e Y_N|\tan\beta}{16 \pi m_{\wt\nu}^2 |M|^2}\times \\
    &\!\left\{\!\left|A_N\right|m_{\chi_1}
    I_{\chi_1}\sin\t_U \sin \theta_V \sin \!\left(\!\varphi_n \!+\! \frac{\varphi_U-\varphi_V}{2}\!\right)\!
    \!+\! m_{\chi_2}I_{\chi_2}\cos\theta_U \cos \theta_V    \sin \!\left(\!\varphi_n
    \!+\frac{\varphi_U+\varphi_V}{2}\!\right)\!\right.\cr
    &\!+\!\left.\left|\!\frac{BY_N}{M}\!\right|\!m_{\chi_1}
    I_{\chi_1}\!\sin\theta_U \sin \theta_V \sin \!\left(\!\!\varphi_n' \!+
    \!\frac{\varphi_U-\varphi_V}{2}\!\right)\!\!
    +\!m_{\chi_2}I_{\chi_2}\cos\theta_U \cos \theta_V    \sin \!\left(\!\!\varphi_n'
    \!+\frac{\varphi_U+\varphi_V}{2}\!\right)\!\right\}\nonumber
\end{align}
where $Q_e=-1$ is the electric charge and
\begin{align*}
    I_{\chi_i}=I_4\left(\frac{m_{\chi_i}^2}{m_{\wt\nu}^2},\frac{m_e^2}{m_{\wt\nu}^2}\right)
    \sim I_4\left(\frac{m_{\chi_i}^2}{m_{\wt\nu}^2},0\right)\;.
\end{align*}
where $I_4(r,s)$ defined in eq.~(\ref{I4}). Here $m_{\tilde\nu}^2$
is the mass-squared of the light sneutrino, $\chi_{1,2}$ are the
two chargino mass eigenstates, $\theta_U$ and $\varphi_U$
($\theta_V$ and $\varphi_V$) are the mixing angle and phase in the
matrix that transforms between the mass eigenbasis and interaction
eigenbasis of the charge $-1(+1$) charginos. A strong suppression
arises from singlet-doublet sneutrino mixing. It is given by
\begin{align}
    -\frac{v_u^* A_N^*}{|M|^2}+\frac{2B^* v_u^* M Y_N^*}{|M|^4}\;.
\end{align}
These terms contribute to $\varphi_n$ and $\varphi_n'$. For
$m_\chi\sim \mu$:
\begin{align}
    \varphi_n\cong \phi_W-\theta \qquad
    \varphi_n'\cong  \phi_W-\theta+\phi_N
\end{align}
However, for a general $m_\chi$, the CP violating phases are more
complicated combinations of $\phi_W$, $\phi_N$ (\ref{phases}) and
$\t$ (\ref{SSB phase}), where the latter is the only CP violating
phase that appears in $\varphi_{U,V}$. The $A_N$ contribution is
analogous to the $A_u$-chargino contribution to the EDM of the up
quark \cite{Ibrahim:1998je}, with the replacement
$N\leftrightarrow d$ and $L\leftrightarrow u$.

In our framework, $\theta_{U,V}={\cal O}(1)$. The result is then
\begin{equation} \label{edmslg}
    |d_e|\approx \frac{e m_e\tan\!\beta }{16\pi m_{\tilde\nu}^2}
    \left|\frac{m_\chi Y_N}{M^2}\right|\left(|A_N|+\left|\frac{B Y_N}{M}\right|\right)
    \lesssim 6 \times 10^{-35}\ e\ \text {cm}\;.
\end{equation}
We use here $A_N/Y_N\sim m_\chi \sim m_{\wt\nu}\sim m_{SUSY}$, $M>
1$ TeV, $\tan\beta<100$, and the bound from leptogenesis for out
of equilibrium decay:
\begin{equation}\label{LG bound}
    \frac{Y_N^2}{M}\lesssim 3\times 10^{-15}\text{ GeV}^{-1}\;.
\end{equation}
The contribution of $B$ is significant only if $M\lesssim10^5$
GeV, when $B$ can assume its natural value, $B\sim M m_{SUSY}$,
since only then it has no restriction from soft leptogenesis
\cite{Grossman:2004dz}. Note that for large $M$, $M\sim 10^9$ GeV,
there is a two loop contribution of the same order, where $B$
generates an $A_L$ term through 1-loop diagrams
\cite{Farzan:2003gn}. For such a high scale for $M$, however, all
contributions are much smaller than the estimate of (\ref{edmslg})
due to the large sterile sneutrino mass. We conclude that the
upper limit (\ref{edmslg}) on the contributions to the EDM of the
electron related to $A_N$ and $B$ is well below the sensitivities
of current (\ref{EDM exp}) and future (\ref{EDM future})
experiments.

Assuming that $\wt N_1$ has comparable mixing with all three
flavors of $\wt \nu$, we can estimate the contribution to $d_\mu$
($d_\tau$) by simply replacing $m_e$ with $m_\mu$ ($m_\tau$)
 in eq.(\ref{edmslg}). The result is:
\begin{align}
   d_\mu &\lesssim  10^{-32}\mbox{e cm}\;,\cr
   d_\tau &\lesssim  2\times10^{-31}\mbox{e cm}\;.
\end{align}
These bounds are far too strong even for the next generation of
experiments (\ref{EDM future}).

\section{Lepton Flavor Violation of Charged Leptons}
\label{LFVsec} The Yukawa interactions of the sterile sneutrino
contribute to $Br(\ell_i\to \ell_j\gamma)$ in two ways: through
the RGE of the slepton masses and through two diagrams similar to
fig.~\ref{EDM fig}, but replacing the incoming (outgoing) electron
with $\ell_i$ ($\ell_j$) and $\wt\phi_u^+$ with Wino. We use here
the formulas of \cite{Hisano:1996qq} to calculate both
contributions, estimate their size, and compare them to
experimental current limits,
\begin{align}\label{Branch exp}
    &Br(\mu\to e\gamma)_{\text{exp}}< 1.2\times 10^{-11} \mbox{ \cite{Brooks:1999pu}}\;,\cr
    &Br(\tau\to e \gamma)_{\text{exp}}<3.7 \times 10^{-7}\mbox{ \cite{Inami:2003}}\;,\cr
    &Br(\tau\to \mu \gamma)_{\text{exp}}<3.1 \times 10^{-7} \mbox{ \cite{Abe:2003sx}}\;,
\end{align}
and near future expected sensitivities,
\begin{align}\label{Branch future}
    &Br(\mu\to e\gamma)_{\text{future}}< 10^{-14}\mbox{\cite{Barkov}},\cr
    &Br(\tau\to e \gamma)_{\text{future}}\lesssim 10^{-8} \mbox{ \cite{Inami:2003}}\;,\cr
    &Br(\tau\to \mu \gamma)_{\text{future}}\lesssim 10^{-8} \mbox{ \cite{Inami:2003}}\;.
\end{align}

We assume universality of the soft breaking terms at the Planck
scale. Only in the framework of universality, flavor changing
couplings come only from $Y_N$ and $A_N$. In a general SSM, there
are arbitrary flavor changing mass-squared terms for sleptons
already at high scale, and their contributions cannot be separated
from the contribution of $Y_N$ and $A_N$. Thus only in models of
universality, there is a correlation between LFV observables and
leptogenesis parameters. We use here renormalization group
equation (RGE) to find the low scale values of the soft breaking
terms.

The leading contribution to the branching ratio comes from gaugino
interactions, and is denoted here by $Br(\ell_i\to
\ell_j\gamma)_{g}$. It depends on the off-diagonal elements of the
doublet slepton mass-squared matrix as follows:
\begin{align}\label{BR gg}
    Br(\ell_i\to \ell_j\gamma)_{g}
    &=\frac{\alpha m_\mu^5}{\tau_\mu^{-1}} \left[\frac{g^2}{(8\pi)^2 m_{\wt\nu}^2}
    f_1\left(\frac{m_\chi^2}{m_{\wt \nu_x}^2}\right) \frac{m_{\wt L_{\mu e}}^2}{m_{SUSY}^2}\right]^2
    \sim \frac {\alpha^3}{G_F^2} \frac{|m^2_{\tilde L ij}|^2}
    {\tilde m^8} \tan^2 \beta\;.
\end{align}
where the last expression is the known approximation
\cite{Casas:2001dv} and
\begin{align*}
    f_1(x)=\frac{2+3x-6x^2+x^3+6\ln x}{6(1-x)^4}\;.
\end{align*}
We take the off-diagonal elements of the slepton mass matrix
${m_{\wt L}}_{ij}^2$ to vanish at the Planck scale, and use RGE to
evolve it to low energy. The relevant one-loop RGE is
\cite{Tobe:2003nx}
\begin{equation}
8\pi^2\mu\frac{d}{d\mu}(m^2_{\tilde L})_{ij}= (m^2_{\tilde
L}+m^2_{\tilde\nu}+m^2_{H_2})(Y_N^\dag Y_N)_{ij} +(A_N^\dagger
A_N)_{ij}\;.
\end{equation}
Assuming that $|A_N|\sim |Y_N m_{SUSY}|$, one gets $(m^2_{\tilde
L})_{ij}/m_{SUSY}^2\sim Y_N^2$\; for $i\neq j$. From soft
leptogenesis we can estimate:
\begin{align}\label{Y limit}
    (Y_N^\dag Y_N)_{11}=\sum_\ell |(Y_N)_{\ell 1}|^2<10^{-8}
    \;\Rightarrow\; |Y_N|_{\ell \,1}\lesssim 10^{-4}\;.
\end{align}
This gives $(m_{\wt L})_{ij}^2/m_{SUSY}^2\lesssim 10^{-8}$ with
$i\neq j$, and consequently
\begin{gather}\label{BR gg estimate}
    Br(\mu\to e\gamma)_{g}\lesssim 10^{-21}\;,\cr
    Br(\tau\to e(\mu) \gamma)_{g}\sim
    \frac{m_\tau^5}{m_\mu^5}Br(\mu\to e \gamma)_{g}\lesssim
    2 \times 10^{-15}\;.
\end{gather}
where we used  $m_{SUSY}\sim 100$ GeV to maximize the possible
effects. Note that this contribution does not involve CP violation.
Comparing this result to (\ref{Branch exp}, \ref{Branch future}), we
conclude that this contribution is unobservably small. The main
reason is that $Br(\mu\to e\gamma)\propto Y_N^4$, and soft
leptogenesis requires a very small Yukawa coupling $Y_N$ (see
(\ref{Y limit})).

The direct contribution of the sneutrino to $\mu \to e \gamma$
involves Yukawa couplings instead of one of the gauge
interactions, and is denoted here as $Br(\mu\to e\gamma)_{Y}$.
Using the expressions of \cite{Hisano:1996qq} and considering, for
simplicity, a two generation model of neutrinos, we get for $\wt
N_1$ contribution
\begin{align}\label{BR gY}
    Br&(\mu\to e \gamma)_{Y}=\frac{\alpha m_\mu^5}{\tau_\mu^{-1}} \left[\frac{g^3 \sin 2\t_V}{2(4\pi)^4 m_{\wt\nu}^4}
    f_{\chi}\right] \frac{m_{\wt L_{\mu e}}^2}{m_{SUSY}^2}
    \frac{1 }{\sqrt 2|M|^2}(\cos\t_{\wt \nu}-\sin \t_{\wt \nu})\cr
    &\times\Re\left[\left(f_{\chi_2} e^{i\varphi_V}-f_{\chi_1}e^{-i\varphi_V}\right)
    \left(-{v_u^* A_N^*{Y_N}_{\mu i}}+\frac{2 B^* v_u^* M \sum_{i,x}({Y_N}_{\mu i}{Y_N}_{i x}^*)}{|M|^2}\right)\right.\cr
    &\left.+\left(f_{\chi_2} e^{-i\varphi_V}-f_{\chi_1}e^{i\varphi_V}\right)
     \left(-{v_u A_N{Y_N}_{\mu i}^*}+\frac{2 B v_u M^*  \sum_{i,x}({Y_N}_{e i}^*{Y_N}_{i x})}{|M|^2}\right)\right]\cr
    &\lesssim\; 4\times 10^{-30}
\end{align}
where $f_{\chi_i}=f_1\left({m_{\chi_i}^2}/{m_{\wt
\nu_x}^2}\right)$, and $\theta_{\wt \nu}$ is the mixing angle
between the light generations.
The CP violating phases are $\theta$ and generalizations of
$\phi_W$ and $\phi_N$ for the case of a two generations model.
Here we approximated $M_2\sim M_1$, $m_{\wt\nu_2}^2\sim m_{\wt
\nu_1}^2$ and $B_{ij}\propto M_{ij}$. Note that (\ref{BR gY}) also
depends on $Y_N^4$, similar to (\ref{BR gg}). This contribution is
much smaller than (\ref{BR gg estimate}), since it is inversely
proportional to $M^2$, which enables us to use the stricter bound
of (\ref{LG bound}) instead of (\ref{Y limit}). We learn that both
contributions are much smaller than present limits and future
sensitivity.

In the SSM+N, the gaugino contribution to $\mu \to e\gamma$ is
almost always dominant in the $\mu - e$ conversion, and hence
there is a relation between the predicted $Br(\mu\to e\gamma)$ and
$R(\mu\to e \mbox{ in Ti})$:
\begin{align}
    \frac{R(\mu\to e \mbox{ in Ti})}{Br(\mu\to e\gamma)}
    &\cong 5\times10^{-3}\;,
\end{align}
where $R(\mu\to e \mbox{ in Ti})\equiv\Gamma(\mu^- Ti\to e^-
Ti^{\mbox{ g.s.}})/\Gamma(\mu^- Ti \mbox{ capture})$ (g.s.~stands
for ground state). The current limit \cite{Dohmen:1993mp} and near
future expected sensitivity \cite{PRISM} are
\begin{align}\label{e mu exp}
    &R(\mu\to e\mbox{ in Ti})_{\text{exp}}< 6.1\times 10^{-13},\cr
    &R(\mu\to e\mbox{ in Ti})_{\text{future}}< 10^{-18}\;.
\end{align}
We learn from (\ref{BR gg estimate}) that our estimate,
\begin{align}
    R(\mu\to e \mbox{ in Ti})\lesssim 6\times10^{-25}\;,
\end{align}
is below the near future expected sensitivity.

\section{Conclusions}
\label{CONCsec} In this work we estimated the contribution to low
energy observables of soft supersymmetry breaking terms within the
range that leads to successful soft leptogenesis. The direct
contribution of $\wt N$ to both the EDM and $Br(\ell_i\to \ell_j
\gamma)$ is inversely proportional to $M^2$. Therefore, the
lighter is $M$, the less suppressed is the singlet sneutrino
contribution. One may think therefore that soft leptogenesis,
which requires $M$ to be lighter than standard leptogenesis, can
have more significant phenomenological consequences. However, both
the electron EDM and the branching ratio of $\ell_i\to \ell_j
\gamma$ strongly depend on $Y_N$:
\begin{align}
    d_e \propto Y_N^2\;,\qquad
    Br(\ell_i\to \ell_j \gamma) \propto Y_N^4\;.
\end{align}
Since the Yukawa couplings of $\wt N_1$ must be small,
$Y_N\lesssim 10^{-4}$, for out of equilibrium decay (\ref {LG
bound}), these contributions are strongly suppressed. We learn
that the contribution of soft supersymmetry breaking terms that
induce soft leptogenesis to low energy observables is much smaller
than other contributions in the SSM.

The analysis here is model independent, in the sense that we do
not use any flavor model for the structure of the Yukawa matrix.
We considered only $\wt N_1$ contributions, and the constraints
from successful soft leptogenesis on $M_1$, ${Y_N}_{1k}$ and soft
breaking terms of $\wt N_1$. The contributions to EDMs and LFV
processes from the heavier singlet sneutrinos are not directly
constrained by leptogenesis, and therefore can be larger.
\footnote{In \cite{Chen:2004xy}, GUT models are used to find $Y_N$
and $M$ for the heavier singlet neutrinos. Then soft leptogenesis
as the dominant source of the baryon asymmetry and $Br(\mu\to
e\gamma)$ within the sensitivity of future experiments can be
simultaneously obtained. However, the dominant contribution to the
LFV interactions there comes from heavier sneutrinos and not from
$\wt N_1$.}

\textbf{Acknowledgments:} I am grateful to Yosef Nir for many
discussions, comments and for careful reading of the manuscript. I
thank Yuval Grossman and Esteban Roulet for their helpful
comments.



\begin{thebibliography}{01}

\bibitem{Minkowski:1977sc}
P.~Minkowski,
Phys.\ Lett.\ B {\bf 67}, 421 (1977).

\bibitem{Gell-Mann:vs}
M.~Gell-Mann, P.~Ramond and R.~Slansky,
Print-80-0576 (CERN).

\bibitem{Yanagida}
T. Yanagida, in {\it Proc. of Workshop on Unified Theory and
Baryon
  Number in the Universe}, eds. O. Sawada and A. Sugamoto (KEK, 1979).

\bibitem{Mohapatra:1979ia}
R.~N.~Mohapatra and G.~Senjanovic,
Phys.\ Rev.\ Lett.\  {\bf 44}, 912 (1980).

\bibitem{Fukugita:1986hr}
M.~Fukugita and T.~Yanagida,
Phys.\ Lett.\ B {\bf 174}, 45 (1986).

\bibitem{Grossman:2003jv}
Y.~Grossman, T.~Kashti, Y.~Nir and E.~Roulet,
Phys.\ Rev.\ Lett.\  {\bf 91}, 251801 (2003)
[arXiv:hep-ph/0307081].

\bibitem{D'Ambrosio:2003wy}
G.~D'Ambrosio, G.~F.~Giudice and M.~Raidal,
Phys.\ Lett.\ B {\bf 575}, 75 (2003) [arXiv:hep-ph/0308031].

\bibitem{Grossman:2004dz}
Y.~Grossman, T.~Kashti, Y.~Nir and E.~Roulet,
to appear in JHEP, arXiv:hep-ph/0407063.

\bibitem{Hisano:1996qq}
J.~Hisano, T.~Moroi, K.~Tobe and M.~Yamaguchi,
Phys.\ Lett.\ B {\bf 391}, 341 (1997) [Erratum-ibid.\ B {\bf 397},
357 (1997)] [arXiv:hep-ph/9605296].

\bibitem{Casas:2001dv}
J.~A.~Casas and A.~Ibarra,
arXiv:hep-ph/0109161.

\bibitem{Masina:2002mv}
I.~Masina and C.~A.~Savoy,
Nucl.\ Phys.\ B {\bf 661}, 365 (2003) [arXiv:hep-ph/0211283].

\bibitem{Masina:2003wt}
I.~Masina,
Nucl.\ Phys.\ B {\bf 671}, 432 (2003) [arXiv:hep-ph/0304299].

\bibitem{Arganda:2004bz}
E.~Arganda, A.~M.~Curiel, M.~J.~Herrero and D.~Temes,
arXiv:hep-ph/0407302.

\bibitem{Masiero:2004js}
A.~Masiero, S.~K.~Vempati and O.~Vives,
arXiv:hep-ph/0407325.

\bibitem{Ibrahim:1998je}
T.~Ibrahim and P.~Nath,
Phys.\ Rev.\ D {\bf 58}, 111301 (1998) [Erratum-ibid.\ D {\bf 60},
099902 (1999)] [arXiv:hep-ph/9807501].

\bibitem{Ellis:2001yz}
J.~R.~Ellis, J.~Hisano, M.~Raidal and Y.~Shimizu,
Phys.\ Lett.\ B {\bf 528}, 86 (2002)
[arXiv:hep-ph/0111324].

\bibitem{Abel:2001vy}
S.~Abel, S.~Khalil and O.~Lebedev,
Nucl.\ Phys.\ B {\bf 606}, 151 (2001) [arXiv:hep-ph/0103320].

\bibitem{Kizukuri:1991mb}
Y.~Kizukuri and N.~Oshimo,
Phys.\ Rev.\ D {\bf 45}, 1806 (1992).

\bibitem{Ellis:2002eh}
J.~R.~Ellis, M.~Raidal and T.~Yanagida,
Phys.\ Lett.\ B {\bf 546}, 228 (2002) [arXiv:hep-ph/0206300].

\bibitem{Ellis:2002xg}
J.~R.~Ellis and M.~Raidal,
Nucl.\ Phys.\ B {\bf 643}, 229 (2002) [arXiv:hep-ph/0206174].

\bibitem{Pascoli:2003uh}
S.~Pascoli, S.~T.~Petcov and W.~Rodejohann,
Phys.\ Rev.\ D {\bf 68}, 093007 (2003) [arXiv:hep-ph/0302054].

\bibitem{Dutta:2003my}
B.~Dutta and R.~N.~Mohapatra,
Phys.\ Rev.\ D {\bf 68}, 113008 (2003) [arXiv:hep-ph/0307163].

\bibitem{Davidson:2003cq}
S.~Davidson,
JHEP {\bf 0303}, 037 (2003)
[arXiv:hep-ph/0302075].

\bibitem{Tobe:2003nx}
K.~Tobe, J.~D.~Wells and T.~Yanagida,
arXiv:hep-ph/0310148.

\bibitem{Raidal:2004vt}
M.~Raidal, A.~Strumia and K.~Turzynski,
arXiv:hep-ph/0408015.

\bibitem{Bernreuther:1990jx}
W.~Bernreuther and M.~Suzuki,
Rev.\ Mod.\ Phys.\  {\bf 63}, 313 (1991) [Erratum-ibid.\  {\bf
64}, 633 (1992)].

\bibitem{Commins:gv}
E.~D.~Commins, S.~B.~Ross, D.~De Mille and B.~C.~Regan,
Phys.\ Rev.\ A {\bf 50}, 2960 (1994).

\bibitem{Deng:2003}
H.~Deng, for the Muon $(g-2)$ Collaboration, Talk presented at
WIN-03, October 2003, Lake Geneva, Wisconsin, USA.

\bibitem{Hagiwara:2002fs}
K.~Hagiwara {\it et al.}  [Particle Data Group Collaboration],
Phys.\ Rev.\ D {\bf 66}, 010001 (2002).

\bibitem{Lamoreaux:2001hb}
S.~K.~Lamoreaux,
arXiv:nucl-ex/0109014.

\bibitem{Aysto:2001zs}
J.~Aysto {\it et al.},
arXiv:hep-ph/0109217.

\bibitem{Ecker:1983dj}
G.~Ecker, W.~Grimus and H.~Neufeld,
Nucl.\ Phys.\ B {\bf 229}, 421 (1983).

\bibitem{Farzan:2003gn}
Y.~Farzan,
Phys.\ Rev.\ D {\bf 69}, 073009 (2004) [arXiv:hep-ph/0310055].

\bibitem{Brooks:1999pu}
M.~L.~Brooks {\it et al.}  [MEGA Collaboration],
Phys.\ Rev.\ Lett.\  {\bf 83}, 1521 (1999)
[arXiv:\eprint{hep-ex/9905013}].


\bibitem{Inami:2003}
K.~Inami, for the Belle Collaboration, Talk presented at WIN-03,
October 2003, Lake Geneva, Wisconsin, USA.

\bibitem{Abe:2003sx}
K.~Abe {\it et al.}  [Belle Collaboration],
Phys.\ Rev.\ Lett.\  {\bf 92}, 171802 (2004)
[arXiv:hep-ex/0310029].

\bibitem{Barkov}
L.~M.~Barkov {\it et al.} [PSI Collaboration], proposal for
experiment, {\url{http://meg.web.psi.ch}}.

\bibitem{Dohmen:1993mp}
C.~Dohmen {\it et al.}  [SINDRUM II Collaboration.],
Phys.\ Lett.\ B {\bf 317}, 631 (1993).

\bibitem{PRISM}
PRISM Collaboration, letter of intent to the J-PARC experiment
(L25), {\url{http://www-ps.kek.jp/jhf-np/LOIlist/pdf/L26.pdf}}.

\bibitem{Chen:2004xy}
M.~C.~Chen and K.~T.~Mahanthappa,
arXiv:hep-ph/0409096.

\end{thebibliography}
\end{document}